\documentclass{article} 
\usepackage{amsmath,amssymb,amsfonts,eucal}
\usepackage{times,mathptm}
\usepackage{wrapfig}
\usepackage{fullpage}
\usepackage[all]{xy}
\newcommand{\remove}[1]{}

\newcommand{\email}[1]{\texttt{#1}}

\newcommand{\C}{{\mathbb C}} 
\newcommand{\F}{{\mathbb F}}
\newcommand{\Z}{{\mathbb Z}}

\newcommand{\tr}{{\textrm{tr}\,}}
\newcommand{\U}{\textrm{U}}

\newcommand{\basis}{B}

\newcommand{\GL}{\textrm{GL}}
\newcommand{\SL}{\textrm{SL}}
\newcommand{\Para}{\textrm{P}}
\newcommand{\PGL}{\textrm{PGL}}
\newcommand{\PSL}{\textrm{PSL}}
\def\s{\scriptstyle}

\newcommand{\qed}{\hfill $\Box$ \medskip}
\newcommand{\polylog}{{\rm polylog}}
\newcommand{\poly}{{\rm poly}}

\newcommand{\Brat}{\mathfrak{B}}

\newcommand{\ket}[1]{\left| #1\right\rangle}
\newcommand{\ve}{\left( \! \begin{array}{c}}
\newcommand{\ctor}{\end{array} \! \right)}
\newcommand{\mat}{\left( \! \begin{array}{rr}}
\newcommand{\rix}{\end{array} \! \right)}
\newcommand{\norm}[1]{\left\| #1 \right\|}

\newcommand{\card}[1]{\left| #1 \right|}
\newcommand{\product}[2]{\left\langle #1, #2 \right\rangle}

\newcommand{\res}[2]{\left.#1\right|_{#2}}

\newcommand{\wreath}{\,\wr\,}

\newtheorem{theorem}{Theorem}

\newtheorem{lemma}{Lemma}
\newtheorem{definition}{Definition}

\begin{document}

\title{\textsc{Generic Quantum Fourier Transforms}}

\author{Cristopher Moore\\
  Department of Computer Science\\
  University of New Mexico\\
  \email{moore@cs.unm.edu}
  \and Daniel Rockmore\\ Department of Mathematics\\
  Dartmouth College\\
  \email{rockmore@cs.dartmouth.edu} \and
  Alexander Russell\\
  Department of Computer Science and Engineering\\
  University of Connecticut\\
  \email{acr@cse.uconn.edu}}

\maketitle

\begin{abstract}
  The \emph{quantum Fourier transform} (QFT) is the principal
  algorithmic tool underlying most efficient quantum algorithms. We
  present a generic framework for the construction of efficient
  quantum circuits for the QFT by ``quantizing'' the \emph{separation
    of variables} technique that has been so successful in the study
  of classical Fourier transform computations.  Specifically, this
  framework applies the existence of computable Bratteli diagrams,
  adapted factorizations, and Gel'fand-Tsetlin bases to offer
  efficient quantum circuits for the QFT over a wide variety a finite
  Abelian and non-Abelian groups, including all group families for
  which efficient QFTs are currently known and many new group
  families. Moreover, the method gives rise to the first
  subexponential-size quantum circuits for the QFT over the linear
  groups $\GL_k(q)$, $\SL_k(q)$, and the finite groups of Lie type,
  for any fixed prime power $q$.

  \remove{Specifically, applying the classical tools of Bratteli
    diagrams, adapted factorizations, and Gel'fand-Tsetlin bases, we
    give the first efficient QFTs for a broad family of groups
    including
  \begin{itemize}
  \item $\GL_k(q)$, $\SL_k(q)$, $\PGL_k(q)$, and $\PSL_k(q)$, for any
    fixed prime $q$,
  \item the symmetric groups $S_n$ (recovering Beals' QFT for these
    groups) and wreath products $G \wreath S_n$ for $|G| = n^{O(1)}$,
    symmetric groups as a special case.
\end{itemize}}
\end{abstract}

\section{Introduction}

Peter Shor's spectacular application of the Fourier transform over the
cyclic group $\Z_n$ in the seminal discovery of an efficient quantum
factoring algorithm \cite{sicomp::Shor1997} has motivated broad
interest in the problem of efficient quantum computation over
arbitrary groups (see, e.g.,
\cite{Beals:1997:QCF,Grigni:2001:QMA,Hallgren:2000:NSR,Hoyer:Efficient,IvanyosMS01,MooreRRS:Hidden,PuschelRB99,Watrous:2001:QAS}).
While this research effort has become quite ramified, two related
themes have emerged: \textsl{(i.)} development of efficient
\emph{quantum Fourier transforms} and \textsl{(ii.)}  development of
efficient quantum algorithms for the \emph{hidden subgroup problem}.
The complexity of these two problems appears to relate intimately to
the group in question: while quantum Fourier transforms and hidden
subgroup problems over Abelian groups are well-understood,
our understanding of these basic problems over non-Abelian groups
remains embarrassingly sporadic.  Aside from their natural appeal,
this line of research been motivated by the direct relationship to the
graph isomorphism problem: an efficient solution to the hidden
subgroup problem over the (non-Abelian) symmetric groups would yield
an efficient quantum algorithm for graph isomorphism.

Over the cyclic group $\Z_n$ the \emph{quantum Fourier transform} refers
to the transformation taking the state
$$
\sum_{z \in \Z_n} f(z) \ket{z} \qquad \text{to the state}\qquad \sum_{\omega \in \Z_n} \hat{f}(\omega) \ket{\omega},
$$
where $f: \Z_n \to \C$ is a function with $\norm{f}_2 = 1$ and
$\hat{f}(\omega) = \sum_{z} f(z)e^{2 \pi i \omega z/n}$ denotes the familiar
discrete Fourier transform at $\omega$. Over an arbitrary finite group
$G$, this analogously refers to the transformation taking the state
$$
\sum_{z \in G} f(z) \ket{z} \qquad \text{to the state}\qquad \sum_{\omega
  \in \hat{G}} \hat{f}(\omega)_{ij} \ket{\omega, i, j},
$$
where $f: G \to \C$, as before, is a function with $\norm{f}_2 = 1$
and $\hat{f}(\omega)_{ij}$ denotes the $i,j$th entry of the Fourier
transform at the representation $\omega$. This is explained further in
Section~\ref{sec:Representation-Theory}.

While there is no known explicit relationship between the quantum
Fourier transform and the hidden subgroup problem over a group $G$,
all known efficient hidden subgroup algorithms rely on an efficient
quantum Fourier transform.  Indeed, it is fair to say that the quantum
Fourier transform is the only known non-trivial quantum algorithmic
paradigm for such problems.

In this article we focus on the 
construction of efficient quantum Fourier transforms. Our research is
motivated by dramatic progress over the last decade in the theory of
efficient \emph{classical} Fourier transforms (see, e.g.,
\cite{Beth:Computational,Clausen:Fast,DiaconisR:Efficient,MaslenR:Separation,Rockmore:FastAbelian}).
These developments have provided a collection of techniques which,
taken together, yield a uniform framework for
the efficient (classical) computation of Fourier transforms over a
wide variety of important families of groups including, for example,
the finite groups of Lie type (properly parametrized) and the
symmetric groups.

We present here an adaptation to the quantum setting of a wide class
of efficient classical Fourier transform algorithms; namely, 
those achieved by the ``separation of variables'' approach. 
This establishes the first generic quantitative relationship between 
efficient classical Fourier transforms and efficient circuits for the 
quantum Fourier transform.  


Specifically, we define a broad class of \emph{polynomially uniform}
groups and show
\begin{theorem}
  \label{thm:main}
  If $G$ is a polynomially uniform group with a subgroup tower $G=G_m >
  \cdots > \{1\}$ with adapted diameter $D$, maximum multiplicity $M$, and
  maximum index $I = \max_i [G_i:G_{i-1}]$, then there is a quantum
  circuit of size $\poly(I \times D \times M \times \log |G|)$ which computes the
  quantum Fourier transform over $G$.
\end{theorem}
This quantifies the complexity of the quantum Fourier transform in
exactly the same fashion as does Corollary 3.1
of~\cite{soda::MaslenR95} in the classical case.  We extend this class
further by showing that it is closed under a certain type of Abelian
extension which may have exponential index.

Together, these results give efficient QFTs --- namely, circuits of $\polylog(|G|)$ size
--- for many families of groups.  These include
\textsl{(i.)} the Clifford groups $\mathbb{CL}_n$; 
\textsl{(ii.)} the symmetric groups, recovering the algorithm of Beals
\cite{Beals:1997:QCF}; 
\textsl{(iii.)} wreath products $G \wreath S_n$ where $|G| = \poly(n)$; 
\textsl{(iv.)} metabelian groups, 
including metacyclic groups such as
the dihedral and affine groups, recovering
the algorithm of H{\o}yer \cite{Hoyer:Efficient}; 
\textsl{(v.)} bounded extensions of Abelian groups such as the generalized quaternions, 
recovering the algorithm of P{\"u}schel et al.  \cite{PuschelRB99}.

Our methods also give the first subexponential size quantum
circuits for the linear groups $\GL_k(q)$, $\SL_k(q)$,
$\PGL_k(q)$, and $\PSL_k(q)$ for fixed prime power $q$, 
various families of finite groups of Lie type,
and the Chevalley and Weyl groups.

The paper is structured as follows.  
Sections~\ref{sec:Representation-Theory} and~\ref{sec:Diagrams}
briefly summarize the representation theory of finite groups, 
the Bratteli diagram, and adapted bases.  We give our algorithms in 
Section~\ref{sec:QFT} along with a list of group families for which
the provide efficient circuits for the QFT.  We conclude with open
problems in Section~\ref{sec:Conclusion}.

\section{Representation theory background}
\label{sec:Representation-Theory}

Fourier analysis over a group $G$ involves expressing arbitrary
functions $f : G \to \C$ as linear combinations of
specific functions on $G$ which reflect the group's structure and
symmetries. If $G$ is Abelian, these
are precisely the {\em characters} of $G$ (the
homomorphisms of $G$ into $\C$). For a general group, they 
are the {\em irreducible matrix elements}, and the Fourier
transform is the change of basis from the basis of delta functions to
the basis of irreducible matrix elements.

In order to  be precise we need  the language of (finite) group representation
theory (see, e.g., Serre \cite{Serre:Linear} for an excellent
introduction). A \emph{representation} $\rho$ of a finite group $G$ is a
homomorphism $\rho:G \to \U(V)$, where $V$ is a (finite) $d_\rho$-dimensional
vector space over $\C$ with an inner product and $\U(V)$ denotes the
group of unitary linear operators on $V$. Fixing an orthonormal basis
for $V$, each $\rho(g)$ may be realized as a $d_\rho \times d_\rho$ unitary matrix.
When a basis has been selected in this way for $V$, we refer to $\rho$ as
a \emph{matrix representation} of $G$; then each of the $d_\rho^2$
functions $\rho_{ij}(g) = [\rho(g)]_{ij}$ is called a {\em matrix element}
(corresponding to $\rho$).  As $\rho$ is a homomorphism, for any $g,h \in G$,
$\rho(gh)=\rho(g) \rho(h)$, 
implying that in general, $\rho_{ij}(gh) = \sum_{k=1}^{d_\rho} \rho_{ik}(g)
\rho_{kj}(h)$.

A matrix representation $\rho$ of $G$ on $V$ is {\em irreducible} if no
subspace (other than the trivial $\{0\}$ subspace and $V$) is mapped
into itself. This is equivalent to the statement that there is no
change of basis that finds a simultaneously block diagonalization (of
given shape) of all $\rho(g)$.  Otherwise the representation is said to
be \emph{reducible}. The irreducible representations will play a role
in the theory analogous to that of the characters of an Abelian group.
Two representations $\rho$ and $\sigma$ are {\em equivalent} if they differ
only by a change of basis, so that for some fixed unitary matrix
$U$,  $\sigma(g) = U^{-1}\sigma(g)U$, for all $g \in
G$. Up to equivalence, a finite group $G$ has a finite number of
irreducible representations equal to the number of its conjugacy
classes.  For a group $G$, we let $\hat{G}$ denote a collection of
representations of $G$ containing exactly one from each isomorphism
class of irreducible representations.

Selecting bases $B$ for the representations of $\hat{G}$ results in a
set of (inequivalent irreducible) matrix representations; when we wish
to be explicit about this selection of bases, we denote such a
collection $\hat{G}_B$. The matrix elements of the matrix
representations $\rho \in \hat{G}_B$ in fact form an orthonormal basis for
the $|G|$-dimensional vector space of complex-valued functions on $G$.
This implies the important relationship between the dimensions of the
irreducible representations of $G$ and $|G|$: $\sum_{\rho \in \hat{G}} d_\rho^2 =
|G|.$ Such a family gives rise to a general definition of Fourier
transform.
\begin{definition} Let $f: G \to \C$; let $\rho : G \to \U(V)$ be a
  matrix representation of $G$. The {\em Fourier transform of $f$ at
    $\rho$}, denoted $\hat{f}(\rho)$, is the matrix
  $$
  \hat{f}(\rho) = \sqrt{\frac{d_\rho}{\card{G}}} \sum_{g \in G} f(g)\rho(g).
  $$
  We typically restrict our attention to $\hat{f}(\rho)$, where $\rho$ is
  irreducible.
\end{definition}

We refer to the collection of matrices $\langle \hat{f}(\rho) \rangle_{\rho \in
  \hat{G}_\basis}$ as the \emph{Fourier transform} of $f$. Thus $f$ is
mapped into $|\hat{G}|$ matrices of varying dimensions. The total
number of entries in these matrices is $\sum d_\rho^2 = \card{G}$, by the
equation mentioned above. The Fourier transform is linear in $f$; with
the constants used above ($\sqrt{d_\rho/\card{G}}$) it is in fact
unitary, taking the $\card{G}$ complex numbers $\langle f(g) \rangle_{g \in G}$ to
$\card{G}$ complex numbers organized into matrices.

For two complex-valued functions $f_1$ and $f_2$ on a group $G$, there
is a natural inner product $\product{f_1}{f_2}$ given by
$\frac{1}{|G|} \sum_g f_1(g)f_2(g)^*$. For any pair of matrix
representations $\rho,\sigma \in \hat{G}_B$, the
corresponding irreducible matrix elements are orthogonal according to
the inner product: let $\rho$ and $\sigma$ be two elements of
$\hat{G}$; then
\begin{equation}
  \label{eqn:orthogonal}
\product{[\rho(\cdot)]_{ij}}{[\sigma(\cdot)]_{kl}} = \begin{cases} 0 &
  \text{if}\;
  \rho \not\cong \sigma\\
  \frac{1}{d_\rho} \delta_{ik}\delta_{jl} & \text{if}\; \rho = \sigma.\end{cases}
\end{equation}

Computation of the Fourier transform (with respect to a given choice
of $\hat{G}$) is equivalent to the change of basis from that of the
point masses to the irreducible matrix elements determined by
$\hat{G}$. This linear map (of the vector space of functions on $G$)
is invertible, with (point-wise) inverse given by the {\em Fourier
  inversion formula}:
$$
f(s) = \sum_{\rho \in \hat{G}} \sqrt{\frac{d_\rho}{|G|}} \tr\Bigl(\rho(s) \hat{f}(\rho)^{-1}\Bigr).
$$

A reducible matrix representation $\rho : G \to U(V)$ may always be
decomposed into irreducible representations; specifically, there is a
basis of $V$ in which each $\rho(g)$ is block diagonal 
where the $i$th block of $\rho(g)$ is precisely $\sigma_i(g)$ for some
irreducible matrix representation $\sigma_i$. In this case we write $\rho = \bigoplus
\sigma_i$. The number of times a given $\sigma \in \hat{G}$ appears in this
decomposition is the \emph{multiplicity} of $\sigma$ in $\rho$. If the
irreducible representation $\sigma_i$ appears with multiplicity $w_i$ in
decomposition of $\rho$, we may write $\rho = \oplus^{w_1} \sigma_1 \ldots \oplus^{w_r} \sigma_r$.

A representation $\rho$ of a group $G$ is also automatically a
representation of any subgroup $H$. We refer to this \emph{restricted}
representation on $H$ as $\res{\rho}{H}$.  Note that in general,
representations that are irreducible over $G$ may be reducible when
restricted to $H$.

\noindent \textbf{Note:} The familiar \emph{Discrete Fourier
  Transform} (DFT) corresponds to the case in which the group is
cyclic. In this case the representations are all one dimensional, and
if $G$ = $\Z_n$, the linear transformation (i.e., the Fourier
transform,) is an order $n$ Vandermonde matrix using the $n$-th roots
of unity.

\pagebreak

\section{Bratteli diagrams, Gel'fand-Tsetlin bases, and adapted diameters}
\label{sec:Diagrams}

The main ingredients for our algorithm are \textsl{(i.)} a tower of
subgroups (or \emph{chain}) which provides a means by which the
Fourier transform on $G$ can be built iteratively as an accumulation
of Fourier transforms on increasingly larger \remove{and larger}
subgroups and \textsl{(ii.)} a natural indexing scheme for the
representations given by paths in the \emph{Bratteli diagram}
corresponding to the group tower and finally \textsl{(iii.)} a
factorization of group elements in terms of a basic set of generators,
which, when judiciously chosen, provide a factorization of the Fourier
transform as a product of structured (direct sums of tensor products)
and sparse matrices. The complexity of a corresponding efficient
Fourier transform which uses these basic ingredients can then be
derived in terms of basic representation-theoretic and combinatorial
data.

\subsection{Bratteli diagrams and Gel'fand-Tsetlin bases}

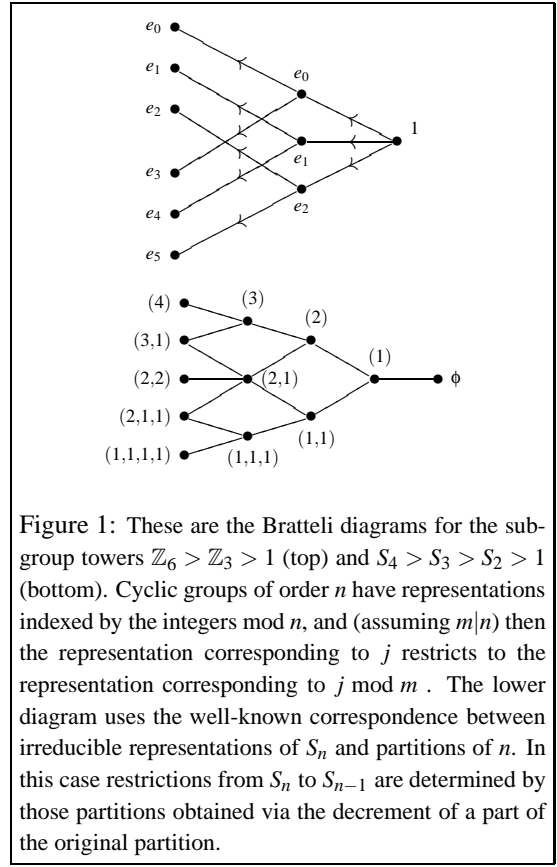
\begin{wrapfigure}{r}{7cm}
\fbox{%
\begin{minipage}{7cm}
  \begin{small}
\begin{displaymath}
{\UseComputerModernTips
\begin{xy}
<1pc,0pc>:
\xycompile{
(0,0)*-={\bullet} ="a0" , (-3,1.5) *-={\bullet} ="b0" , 
(-3,0)*-={\bullet} ="b1" , (-3,-1.5) *-={\bullet} ="b2" ,
(-7,3.6) *-={\bullet} = "c0", (-7,2.3) *-={\bullet} = "c1",
(-7,1.0) *-={\bullet} = "c2", (-7,-1.0) *-={\bullet} = "c3", 
(-7,-2.3) *-={\bullet} = "c4", (-7,-3.6) *-={\bullet} = "c5", 
 \ar @{-}|@{>} "a0";"b0"
 \ar @{-}|@{>} "a0";"b1"
 \ar @{-}|@{>} "a0";"b2"
 \ar @{-}|@{>} "b0";"c0"
 \ar @{-}|@{>} "b1";"c1"
 \ar @{-}|@{>} "b2";"c2"
 \ar @{-}|@{>} "b0";"c3"
 \ar @{-}|@{>} "b1";"c4"
 \ar @{-}|@{>} "b2";"c5"
%
%
\POS"a0"!{+R*+!LD{\s 1}} 
\POS"b0"!{+U*+!D{\s e_0}}
\POS"b1"!{+D*+!U{\s e_1}}
\POS"b2"!{+D*+!U{\s e_2}}
\POS"c0"!{+L*+!R{\s e_0}}
\POS"c1"!{+L*+!R{\s e_1}}
\POS"c2"!{+L*+!R{\s e_2}}
\POS"c3"!{+L*+!R{\s e_3}}
\POS"c4"!{+L*+!R{\s e_4}}
\POS"c5"!{+L*+!R{\s e_5}}
}\end{xy} }
\end{displaymath}
\begin{displaymath}
{\UseComputerModernTips
\begin{xy}
<1.0pc,0pc>:<0pc,0.6pc>::
\xycompile{
(0,0)*-={\bullet} ="a" , (-2,0) *-={\bullet} ="b" , 
(-4,2)*-={\bullet} ="c" , (-4,-2) *-={\bullet} ="d" ,
(-6,3)*-={\bullet} ="e" , (-6,0) *-={\bullet} ="f" ,
(-6,-3) *-={\bullet} ="g" , (-8,4) *-={\bullet} ="h" ,
(-8,2) *-={\bullet} ="i" , (-8,0) *-={\bullet} ="j" ,
(-8,-2) *-={\bullet} ="k" , (-8,-4) *-={\bullet}  ="l"
 \ar @{-} "a";"b"
 \ar @{-} "b";"c"
 \ar @{-} "c";"e"
 \ar @{-} "e";"h"
 \ar @{-} "b";"d"
 \ar @{-} "d";"g"
 \ar @{-} "g";"l"
 \ar @{-} "c";"f"
 \ar @{-} "f";"i"
 \ar @{-} "d";"f"
 \ar @{-} "f";"k"
 \ar @{-} "e";"i"
 \ar @{-} "f";"j"
 \ar @{-} "g";"k"
\POS"a"!{+R*+!L{\s \phi}} 
\POS"b"!{+RU*+!D{\s (1)}} 
\POS"c"!{+RU*+!D{\s (2)}}
\POS"d"!{+RD*+!U{\s (1,1)}} 
\POS"e"!{+RU*+!D{\s (3)}} 
\POS"f"!{+R*+!L{\s (2,1)}} 
\POS"g"!{+RD*+!U{\s (1,1,1)}}
\POS"h"!{+L*+!R{\s (4)}} 
\POS"i"!{+L*+!R{\s (3,1)}} 
\POS"j"!{+L*+!R{\s (2,2)}} 
\POS"k"!{+L*+!R{\s (2,1,1)}}
\POS"l"!{+L*+!R{\s (1,1,1,1)}}
 }\end{xy} }
\end{displaymath}
\caption{%
  {\small These are the Bratteli diagrams for the subgroup
    towers $\Z_6 > \Z_3 > 1$ (top) and $S_4 > S_3 > S_2 > 1$ (bottom).
    Cyclic groups of order $n$ have representations indexed by the
    integers mod $n$, and (assuming $m|n$) then the representation
    corresponding to $j$ restricts to the representation corresponding
    to $j \bmod m$ . The lower diagram uses the well-known
    correspondence between irreducible representations of $S_n$ and
    partitions of $n$. In this case restrictions from $S_n$ to
    $S_{n-1}$ are determined by those partitions obtained via the
    decrement of a part of the original partition.
  }
  \label{fig:diagrams}}
\end{small}
\end{minipage}}
\end{wrapfigure}

Much of Abelian Fourier analysis is simplified by the fact that in
this case the dual (that is, the set of characters $\chi : G \to \C$) also
forms a group isomorphic to the original group; furthermore, in this
isomorphism lies a natural correspondence providing an indexing of the
irreducible representations (i.e., matrix elements).  However, in the
general case there is no immediate indexing scheme for the dual
$\hat{G}$ and the landscape is further complicated by the absence of a
canonical basis for the now multidimensional representations. Indeed,
for the goal of efficient Fourier analysis, not all bases are created
alike!  In particular, a fairly general methodology for the
construction of group FFTs, the "separation of variables" approach
\cite{soda::MaslenR95,MaslenR:Separation} relies on the use of
\emph{Gel'fand-Tsetlin} or \emph{adapted} bases for efficient
computation.  These bases allow for a Fourier transform on $G$ to be
built from Fourier transforms on subgroups, a general technique whose
efficiency improves as it is used through a tower of subgroups. This
is in fact the main idea in the famous ``Cooley-Tukey"
(decimation-in-time) FFT.

A crucial ingredient of the general separation of variables approach
is the incorporation of an indexing scheme that permits the
computational to be organized efficiently. The same {\em Bratteli
  diagram} formalism is key to both the organization and manipulation
of the calculation for a quantum FFT; we present it below.

Given a finite group $G$ and let 
\[
G = G_m > G_{m-1} > 
\cdots >  G_1 >  G_0 = \{1\}
\]
be a tower of subgroups of length $m$ for $G$. The corresponding
\emph{Bratteli diagram}, denoted $\Brat$, is a leveled
directed multigraph whose nodes of level $i = 0,\dots, m$ are in
one-to-one correspondence with the (inequivalent) irreducible
representations of $G_i$. For convenience, we refer to vertices in the
diagram by the representation with which they are associated. The
number of edges from an irreducible representation $\eta$ of $G_i$ to $\rho$
of $G_{i+1}$ is equal to the \emph{multiplicity} of $\eta$ in the
restriction of $\rho$ to $G_i$.  Since there is a unique irreducible
representation of the trivial group, a Bratteli diagram for a given
tower is in fact a rooted tree.

Thus, the edges out of a node $\eta$ of $\hat{G_i}$ represent a complete
set of orthogonal embeddings of the corresponding representation space
into the representations of $G_{i+1}$ and conversely, the edges
entering a given representation $\rho : G_{i+1} \to U(V_\rho)$ of $G_{i+1}$
index a set of mutually orthogonal subspaces of $V_\rho$ whose direct sum
represents the decomposition of $V_\rho$ under the (restricted) action of
$G_i$.  Thus, the paths from the root node to a vertex $\rho : G_{i} \to
U(V_\rho)$ index a basis of $V_\rho$ with the following property: for any
$G_j < G_i$, there is a partition of the basis vectors into subsets,
each of which spans an irreducible $G_j$-invariant subspace, so that
the associated matrix representation is block diagonal according to
this partition when restricted to $G_j$ and, moreover, that blocks for
equivalent irreducible representations are actually equal. Such bases
are said to be (subgroup-)adapted or {\em Gel'fand-Tsetlin}.
Consequently, the number of paths to a node $\eta$ is equal to $d_\eta$, and
pairs of path with common endpoint $\eta$ index an irreducible matrix
element of $\eta$.

The block diagonal nature of the restriction (combined with the fact
that blocks corresponding to equivalent representations are actually
equal) allows the Fourier transform on $G= G_{m}$ to be expressed as a
sum of Fourier transforms on $G_{m-1}$, each translated from a
distinct coset: specifically, if $T \subset G$ is a transversal, i.e.\ a 
set of representatives for the left cosets of $G_{m-1}$ in $G_m$, 
we define $f_\alpha : G_{m-1} \to \C$ by $f_\alpha(x) = f(\alpha x)$.  
Then
\begin{equation}
\hat{f}(\rho) = \sum_{\alpha \in T}
\rho(\alpha) \sum_{x\in G_{m-1}} \rho(x)f (\alpha x) = 
\sum_{\alpha \in T} \rho(\alpha) \cdot \hat{f_\alpha}(\res{\rho}{G_{m-1}}).
\label{eq:basic-rec}
\end{equation}

\subsection{Strong generating sets and adapted diameters}
\label{sec:stronggen}

Adapted representations are only part of the story for the
construction of efficient Fourier transform algorithms.  
In general, $\rho(\alpha)$ of Equation~(\ref{eq:basic-rec}), the ``twiddle factor", could be 
an arbitrary matrix of exponential size, so implementing it 
in \eqref{eq:basic-rec} could be costly. Luckily, under fairly mild
assumptions, the matrices $\rho(\alpha)$ can be factored into $\polylog(|G|)$ 
sparse, highly structured matrices, and can therefore be implemented
with $\polylog(|G|)$ elementary quantum operators.  

We say that $S$ is a {\em strong generating set} for the tower of subgroups
$\{G_i\}$ if $S \cap G_i$ generates $G_i$.  Say that we have chosen a
transversal $T_i$ for each $i$ indexing the cosets of $G_{i-1}$ in $G_i$.  
Now define $D_i = \min \{ \ell > 0: \cup_{j \leq \ell} (S \cap G_i)^j \supseteq T_i \}$, 
and define the {\em adapted diameter} $D = \sum_i D_i$.  Then clearly any 
group element can be factored as a series of coset representatives, 
which in turn can be factored as a total of at most $D$ elements of $S$.  

Of course, to perform the QFT efficiently we would like $\rho(\gamma)$ 
to have a simple form for each $\gamma \in S$.  
Given a subgroup $K < G$, recall that the {\em centralizer} of $K$ 
is the subgroup $Z(K) = \{g \in G: gk=kg \mbox{ for all } k \in K\}$.
The following is implicit in the oft-cited lemma of Schur:

\begin{lemma} (Schur, \cite[Lemma 5.1]{soda::MaslenR95}) Let $K < G$, let 
$\gamma \in Z(K)$, and let $\rho$ be a $K$-adapted representation of $G$. 
Suppose that $\res{\rho}{K} = \oplus^{m_1}\eta_1\cdots \oplus^{m_r} \eta_r$. Then $\rho(\gamma)$ has the form
\begin{equation}
\label{eq:sparsefactors}
(\GL_{m_1}(\C)\otimes I_{d_1})\oplus\cdots \oplus (\GL_{m_r}(\C)\otimes I_{d_r})
\end{equation}
where $I_k$ is the $k \times k$ identity matrix and $d_i = d_{\eta_i}$.
\end{lemma}

\noindent
Since any unitary operator in $\GL_m(\C)$ can be carried out with 
$\poly(m)$ elementary quantum gates~\cite{barencoetal}, 
and since we can condition on the $\eta_i$ to find out which subspace of
$\rho$ we are in, we can write $\rho(\gamma)$ as a 
series of $\poly(M)$ elementary quantum operations 
where $M = \max_i m_i$ in~\eqref{eq:sparsefactors}.
Therefore, the total number of elementary quantum operators we need to 
implement $\rho(\alpha)$ is then $D \times \poly(M)$.

Moreover, if $\gamma$ is itself in a subgroup $H > K$, and $\rho$ is
adapted to both $H$ and $K$, then $\rho(a)$ also possesses the block
structure corresponding to $\res{\rho}{H}$.  This places an upper bound 
on $M$ of the maximum multiplicity with which representations of $K$
appear in restrictions of representations of $H$.  Thus we can minimize 
$M$ by choosing generators
$\gamma$ inside subgroups as low on the tower as possible, 
which centralize subgroups as high on the tower as possible.

For instance, in the symmetric group $S_n$ we take the tower to be
$S_n > S_{n-1} > \cdots > \{1\}$, where $S_i$ fixes all elements 
greater than $i$.  Let $S$ be the set of 
pairwise adjacent transpositions $(j,j+1)$; each of these is contained
in $S_{j+1}$ and centralizes $S_{j-1}$.  The maximum multiplicity
with which a representation of $S_{j-1}$ appears in a representation of
$S_{j+1}$ is 2, corresponding to the two orders in which we can remove 
two cells from a Young diagram.
Since the adapted diameter is easily seen to be $O(n^2)$, this means
that the $\rho(\alpha)$ can be carried out in $O(n^2) = \polylog(|S_n|)$ 
elementary quantum operations~\cite{Beals:1997:QCF}.  We will see that 
a similar situation obtains for a large class of groups.


\section{Efficient quantum Fourier transforms}
\label{sec:QFT}

We describe our algorithm in this section. The algorithm performs the
Fourier transform inductively on the tower of subgroups, using the
structure of the Bratteli diagram to construct the transform at each 
level from the transform at the previous level.

Recall that for each level of our tower of subgroups $G = G_m > G_{m-1} 
> \cdots > G_0 = \{1\}$ we have chosen a transversal $T_i$ for the left
cosets of $G_{i-1}$ in $G_i$.  At the beginning of the computation, we
represent each group element $g$ as a product 
$\alpha = \alpha_m \cdots \alpha_1$ where $\alpha_i \in T_i$.  
This string becomes
shorter as we work our way up the tower, and after having performed
the Fourier transform for $G_i$ the remaining string $\alpha = \alpha_m \cdots
\alpha_{i+1}$ indexes the coset of $G_i$ in $G$ in which $g$ lies.

At the end of the computation, we have a pair of paths in the Bratteli
diagram, $s=s_1 \cdots s_m$ and $t=t_1 \cdots t_m$, which index the
rows and columns of the representations $\rho$ of $G$.  These paths
begin empty and grow as we work our way up the tower; after
having performed the Fourier transform for $G_i$, the 
paths $p=p_1 \cdots p_i$ and $q=q_1 \cdots q_i$ of length $i$
index the rows and columns of representations $\sigma$ of $G_i$.

With a compact encoding, one could store $\alpha$ in the same registers
as $s$ and $t$, at each step replacing a coset representative
$\alpha_i$ with a pair of edges $s_i, t_i$.  However, our algorithm is 
simpler to describe if we double the number of qubits and store $\alpha$ 
and $s,t$ in separate registers.  
Padding out $\alpha$, $s$, and $t$ to length $m$ with zeroes,
our computational basis consists of unit vectors of the form
\[ 
\ket{\alpha} \ket{s,t} =
\ket{\alpha_m \cdots \alpha_{i+1} \,0^i} 
\otimes \ket{s_1 \cdots s_i \,0^{m-i}, s_1 \cdots s_i \,0^{m-i}} 
\enspace .
\]
Keep in mind the basis $\{\ket{s,t}\}$, where $s$ and $t$ have 
length $i$ and end in the same representation, is just a 
permutation of our adapted Gel'fand-Tsetlin basis $\{\ket{\sigma,j,k}\}$ 
for $\hat{G}_i$, where $\sigma$ ranges over the 
representations of $G_i$ and $1 \leq j,k \leq d_\sigma$ index 
its rows and columns.  Therefore, we will sometimes abuse 
notation by writing $\hat{f}(s,t)$ and $\hat{f}(\sigma)_{j,k}$
for the Fourier transform over $G_i$ indexed in these two different ways.

Each stage of the algorithm consists of calculating the Fourier
transform over $G_{i+1}$ from that over $G_i$. 
By induction it suffices to consider the last stage, where we go from 
$H=G_{m-1}$ to $G=G_m$.
Specifically, choose a transversal $T$ of $H$ in $G$ such that
every $g \in G$ can be written $\alpha h$ where $\alpha \in T$
and $h \in H$.  For each $\alpha \in T$, define a function 
$f_\alpha$ on $H$ as $f_\alpha(h) = f(\alpha h)$; this
is the restriction of $f$ to the coset $\alpha H$, shifted into $H$.  

After having performed the Fourier transform on $H$, our state
will be
\begin{equation}
\label{eq:stateh}
\sum_{\alpha \in T} \ket{\alpha} \;\otimes
\sum_{s,t\; \text{of length}\; m-1}
\hat{f}_\alpha(s,t) \ket{s,t} 
\;\;=\;\;
\sum_{\alpha \in T} \ket{\alpha} \;\otimes
\sum_{(\sigma,j,k) \in \hat{H}}
\hat{f}_\alpha(\sigma)_{j,k} \ket{\sigma,j,k} 
\enspace .
\end{equation}
Our goal is to transform this state into the Fourier basis of $G$, namely
\begin{equation}
\label{eq:stateg}
\ket{0} \;\otimes 
\sum_{s,t\; \text{of length}\;m}
\hat{f}(s,t) \ket{s,t} 
\;\;=\;\;
\ket{0} \;\otimes
\sum_{(\rho,j,k) \in \hat{G}}
\hat{f}(\rho)_{j,k} \ket{\rho,j,k} 
\enspace .
\end{equation}
where $\ket{0}$ occupies the register that held the coset
representative $\alpha$ before.

This transformation is greatly simplified by the following two
observations, which are common to nearly every algorithm for the FFT.
First, as described in Equation~(\ref{eq:basic-rec}) above, $\hat{f}$
can be written as a sum over contributions from $f$'s values on each
coset $\alpha H$, giving
\begin{equation}
\label{eq:sumovercosets}
\hat{f}(\rho) = \sum_{\alpha \in T} \rho(\alpha) \cdot \hat{f}_\alpha(\rho)
\enspace .
\end{equation}
Since $f_\alpha$ has support only in $H$, the matrix $\hat{f}_\alpha(\rho)$ is 
a direct sum of sub-matrices of the form  $\hat{f}_\alpha(\sigma)$, 
summed over the $\sigma$ appearing in $\rho$.  In the quantum setting we
accomplish this via 
an {\em embedding} operation which reverses the restriction to $H$,
\begin{equation}
\label{eq:embedding}
 \ket{\sigma} \to 
   \sum_{\rho: \,\sigma\; \text{appears in}\; \res{\rho}{H}} 
   A_{\sigma,\rho} \ket{\rho}
\end{equation}
where this ``scale factor'' is
\[ A_{\sigma,\rho} = \sqrt{\frac{|H|}{|G|} \frac{d_\rho}{d_\sigma}} 
\enspace . \]
(Note that $\sum_{\rho} |A_{\sigma,\rho}|^2 = 1$.)

Thus the algorithm consists of \textsl{(i.)} embedding the $\sigma$ in the
appropriate $\rho$, \textsl{(ii.)} applying the ``twiddle factor" $\rho(\alpha)$,
and \textsl{(iii.)} summing over the cosets.  However, in general, doing these
things efficiently is no simple matter.  First, a given $\sigma$ might
appear in a given $\rho$ with an arbitrary change of basis; the twiddle
$\rho(\alpha)$ could be an arbitrary unitary matrix of exponential size; and
summing over an exponential number of cosets will take exponential
time unless parallelized in some way.

It is here that the Bratteli diagram proves to be extremely helpful.  
It allows us to implement the twiddle factors $\rho(\alpha)$ efficiently 
when coupled with a strong generating set as discussed in 
Section~\ref{sec:stronggen}  by providing an adapted basis.  
It simplifies the embedding operation as well: 
first note that $\hat{f}_\alpha(s,t)$ is nonzero only when
$s$ and $t$ end in the same representation $\sigma$ of $G_t$, 
i.e.\ in the same vertex of the diagram.   Moreover,
recall that the Bratteli diagram indexes an adapted basis in which 
$\res{\rho}{H}$ is block-diagonal with the $\sigma_j$ as its blocks.  
This means that the $\sigma$ appear in the $\rho$ in an extremely 
simple way: namely, where $s$ and $t$ are extended by appending
the same edge $e$ to both.  

Let adopt some notation.  
Given a path $s$ in the Bratteli diagram of length $m-1$ or $m$, 
denote the representation in which it ends by $\sigma[s]$ or $\rho[s]$ 
respectively, and if $s = s_1 \cdots s_{m-1}$, denote
$s_1 \cdots s_{m-1} e$ as $se$.  We will index the edges of 
each vertex $\{1,\ldots,k\}$ where it has out-degree $k$.  
It will be convenient to carry out this embedding only if the register 
containing the coset representative is zero, and leave other basis 
vectors in $(T \cup \{0\}) \otimes \hat{H}$ fixed.
Then~\eqref{eq:embedding} becomes
\begin{equation}
\label{eq:ourembedding}
U: \left\{ \begin{array}{l}
\ket{0} \ket{s,t} \to \ket{0} \sum_e A_{\sigma[s],\rho[se]} \ket{se,te} \\
\ket{\alpha} \ket{s,t} \to \ket{\alpha} \ket{s,t} \;\text{for all}\; \alpha \in T
\end{array} \right.
\end{equation}
where the sum is over all outgoing edges $e$ of $\sigma[s]=\sigma[s]$.  

Note that we have not defined $U$ on the entire space; in particular,
since we are moving probability from $\hat{H}$ to $\hat{G}$, basis vectors
$\ket{0} \ket{se,te} \in (T \cup \{0\}) \otimes \hat{G}$ cannot stay fixed.  
As we will see below, it does not matter precisely how $U$ behaves on the 
rest of the state space, as long as its behavior on $\hat{H}$ is as described 
in~\eqref{eq:ourembedding}. 
This can be accomplished simply by putting the $m$'th registers 
of $s$ and $t$ in the superposition 
$\sum_e A_{\sigma[s],\rho[se]} \ket{e} \otimes \ket{e}$, and for a large class
of extensions we can prepare this superposition efficiently.

\subsection{Extensions of subexponential index}

In this section we generalize Beals' QFT for the symmetric 
group~\cite{Beals:1997:QCF} to a large class of groups.  
First we show that the Fourier transform can be extended from $H$ to $G$, 
modulo some reasonable uniformity conditions on $G$.

\begin{definition}  For a group $G$ and a tower of subgroups $G_i$,
let $\Brat$ be the corresponding Bratteli diagram, let $T_i$ be a
set of coset representatives at each level, and let $S$ be
a strong set of generators for $G$.  
Then we say that $G$ is
{\em polynomially uniform} (with respect to $\{G_i\}$, $\Brat$, $\{T_i\}$, and $S$)
if the following functions are computable
by a classical algorithm in $\polylog(|G|)$ time:
\begin{enumerate}
\item Given two paths $s,t$ in $\Brat$, whether $\rho[s]=\rho[t]$;
\item Given a path $s$ in $\Brat$, the dimension and
the out-degree of $\rho[s]$;
\item Given a coset representative $\alpha_i \in T_i$, a factorization
of $\alpha$ as a word of $\polylog(|G|)$ length in $(S \cap G_i)^*$.
\end{enumerate}
\end{definition}

\begin{lemma}
\label{lem:polyindex}
If $G$ is polynomially uniform with respect to a tower of subgroups 
where $G=G_m$ and $H = G_{m-1}$ and a strong generating set $S$
with adapted diameter $D$ and maximum multiplicity $M$, 
then the Fourier transform of $G$ can obtained
from the state~\eqref{eq:stateh} using
$\poly([G:H] \times D \times M \times \log |G|)$ 
elementary quantum operations.
\end{lemma}

\noindent
{\em Proof.}  First, to carry out the embedding transformation $U$, we use the 
classical algorithm to compute the list of edges $e$ and $d_{\rho[se]}$ 
conditional on $s$, and thus compute the $A_{\sigma,\rho}$ (say, to
$n$ digits in $\poly(n)$ time).  Note that $\sigma$ appears in at most 
$[G:H]$ many $\rho$.
We then carry out a series of $[G:H]$ conditional rotations, 
each of which rotates the appropriate amplitude
from $\ket{0} \ket{s,t}$ to $\ket{0} \ket{se,te}$.  Thus $U$, and therefore
$U^{-1}$, can be carried out in $O([G:H])$ quantum operations.  

To apply the twiddle factor and sum over the cosets as 
in~\eqref{eq:sumovercosets},
we use a technique of Beals~\cite{Beals:1997:QCF} and 
carry out the following for-loop.  For each $\alpha \in T$, we do the 
following three things: left multiply $\hat{f}(\rho)$ by
$\rho(\alpha)^{-1}$; add $\hat{f}_\alpha(\rho)$ to $\hat{f}(\rho)$; and
left multiply $\hat{f}(\rho)$ by $\rho(\alpha)$.  This loop clearly produces
$\sum_{\alpha \in T} \rho(\alpha) \cdot \hat{f}(\rho)$, so we just 
need to show that each of these three steps can be carried out
efficiently.

Recall that $\hat{f}(\rho)$ is given in the 
$\ket{s,t}$ basis, where $s$ and $t$ index the row and column
of $\rho$ respectively.  To left multiply $\hat{f}(\rho)$
by $\rho(\alpha)$, we apply $\rho(\alpha)$ to the $s$ register
and leave the $t$ register unchanged.  Since $G$ is polynomially 
uniform, a classical algorithm can factor $\alpha$ 
as the product of $D$ generators $\gamma_i \in S$, 
and provide a factorization of each $\rho(\gamma_i)$ as the product 
of $\poly(M)$ many elementary quantum operations, in 
$\polylog(|G|)$ time.  This implements 
$\rho(\alpha)$ and $\rho(\alpha)^{-1}$ in
$D \times \poly(M) + \polylog(|G|)$ operations.

The step ``add $\hat{f}_\alpha(\rho)$ to $\hat{f}(\rho)$'' is slightly more
mysterious, and indeed it does not even sound unitary at first.  
However, as Beals points out, at each point in the loop we are 
adding $\hat{f}_\alpha(\rho)$, which is the Fourier transform of a
function with support only on $H$, to 
$\sum_{\beta < \alpha} \rho(\alpha^{-1} \beta) \hat{f}_\beta(\rho)$, 
which is the Fourier transform of a function with support only
{\em outside} $H$.  Thus these two states are orthogonal, and 
adding two orthogonal vectors can be done unitarily
by rotating one vector into the other 
while fixing the subspace perpendicular to both.  Let 
$V_\alpha$ be the operation that exchanges $\ket{\alpha}\ket{s,t}$
with $\ket{0}\ket{s,t}$ and leaves $\ket{\beta}\ket{s,t}$ fixed for all
$\beta \leq \alpha, 0$; then Beals showed that this step can be written
$U^{-1} V_\alpha U$ where $U$ is the embedding operator defined 
in~\eqref{eq:ourembedding}.  We showed earlier that $U$ can be
carried out in $O([G:H])$ quantum operations, and $V$ is a 
simply a Boolean operation on the $\alpha$ register.  Finally, 
the for-loop runs $|T|=[G:H]$ times, so we're done.
\qed

\noindent
{\em Proof of Theorem~\ref{thm:main}.}  
This follows immediately from the fact that the depth of the Bratteli diagram 
is at most $\log |G|$.
\qed

As noted above, for many groups, the maximum index $I = \max_i [G_i:G_{i-1}]$,
the adapted diameter $D$, and the maximum multiplicity $M$ are all 
$\polylog(|G|)$.  In this case, Theorem~\ref{thm:main} gives circuits for the QFT of
$\polylog(|G|)$ size.  This includes the following three families of groups:

\medskip \noindent \textbf{The symmetric groups $S_n$.} As stated above, 
we take the tower $S_n > S_{n-1} > \cdots > \{1\}$ where $S_i$
fixes all elements greater than $i$.  The maximum index is then $n = o(\log |S_n|)$.
The generators are the adjacent transpositions; the adapted diameter is $O(n^2)$ 
and the maximum multiplicity is 2.  The adapted basis is precisely the Young
orthogonal basis.

\medskip \noindent \textbf{Wreath products $G = H \wr S_n$ for $H$ of
  size $\poly(n)$.}  These groups arise naturally as automorphism groups of
graphs obtained by composition \cite{Harary:Graph}. As in~\cite{Rockmore:FastWreath} 
the tower is
$$
H \wr S_n \,>\, H \times (H \wreath S_{n-1}) \,>\, H \wreath S_{n-1} \,>\,
\cdots \,>\, \{1\} \enspace .
$$
The maximum index is $\max(n,|H|)$, the generators are the adjacent
transpositions and an arbitrary set of $\log |H|$ generators for each
factor of $H$, the adapted diameter is $O(n^2 \log |H|)$, and the
maximum multiplicity is $O(|H|)$.  Then note that $|H| =
\polylog(|G|)$.  See \cite{soda::MaslenR95} for details and
\cite{Kerber:Representations} for discussion on wreath products. 

\medskip \noindent \textbf{The Clifford groups.} The Clifford groups $\mathbb{CL}_n$ 
are generated by $x_1, \ldots, x_n$ where $x_i^2 = 1$ and $x_i x_j = -x_i x_j$
for all $i \neq j$ \cite{Simon:Representations}.  We take the tower 
$\mathbb{CL}_n > \mathbb{CL}_{n-1} > \cdots > \{1\}$ which has maximum 
index $2$, and the generators $\bigl\{ \{x_1\}, \{x_1 x_2\},
\ldots, \{x_{n-1} x_{n}\} \bigr\}$.  The adapted diameter is $O(n)$, and since 
each $x_i x_{i+1}$ centralizes $\mathbb{CL}_{i-1}$, the maximum multiplicity is $4$.

\medskip
In addition to giving $\polylog(|G|)$-size circuits for these groups, this technique
also gives the first subexponential-size circuits for the following classical groups:

\begin{wrapfigure}{r}{2cm}
      \label{fig:parabolic}
  \fbox{%
    \begin{minipage}{2cm}
      \begin{small}
        \begin{center}
          \[
          \left(\begin{array}{c|c}
              A & v\\
              \hline\\
              0\dots 0 & c \end{array}\right)
          \]      
        \end{center}
        \caption{$P_k$}
      \end{small}
  \end{minipage}}
\end{wrapfigure}

\medskip \noindent \textbf{The linear groups $\GL_n(q)$, $\SL_n(q)$,
  $\PGL_n(q)$, and $\PSL_n(q)$; the finite groups of Lie type; the
  Chevalley and Weyl groups.} The case of $\GL_n(q)$ is emblematic of
all these families. We have a natural tower:
$$
\GL_n(q) \,>\, \Para_n(q) \,>\, \GL_{n-1}(q) \times \GL_1(q) \,>\, \GL_{k-1}(q) \,>\, \{1\}
\enspace .
$$
Here $\Para_k(q)$ is the so-called maximal parabolic subgroup of
the form shown in Figure 2, where $A \in \GL_{k-1}(q), v\in
\F_q^{k-1}$, and $c\in \F_q^\times$.  
Our generators are block-diagonal with an arbitrary element of 
$\GL_2(q)$ in the $i,i-1$ block and all other diagonal elements
equal to 1.  The adapted diameter
is $O(n^2)$, the maximum index is $q^{n-1}$, and the 
maximum multiplicity is $q^{O(n)}$.  Analogous factorizations
arise in the case of the finite groups of Lie type as well as
the finite unitary groups~\cite{MaslenR:Separation}.  

Theorem~\ref{thm:main} then implies a quantum circuit of size $q^{O(n)}$ 
for the QFT over these groups.  Since $|G|=O(q^{n^2})$ we can write this as 
$|G|^{O(1/n)}$, which is $\exp\bigl(O(\sqrt{\log |G|})\bigr)$ if $q$ is fixed.  
Note that the best-known classical algorithm for these 
groups~\cite{soda::MaslenR95} has complexity $|G| \,q^{\Theta(n)} 
= G^{1+\Theta(1/n)}$;  therefore, we argue that this quantum speedup 
is the most we could expect relative to the existing classical algorithm.  
Note, for instance, that for the group families above for which we obtain
circuits of size $\polylog(|G|)$, there are classical algorithms of complexity
$|G| \,\polylog(|G|)$.  In both cases it appears that the natural quantum speedup
is to remove a factor of $|G|$ (modulo polylogarithmic terms).

\subsection{Extensions of exponential index and Coppersmith-type circuits}

The reader familiar with Coppersmith's circuit~\cite{Coppersmith:Approximate} for the QFT 
over $G=\Z_{2^n}$, where $H = \Z_{2^{n-1}}$, will recall that the 
Hadamard gate embeds a character $\sigma \in \hat{H}$ 
in two characters $\rho \in \hat{G}$, applies part of the 
twiddle factor, and sums over the two cosets of $H$, 
all in one operation.  This is in contrast to Beals' technique, 
which sums over the cosets serially.  Indeed, if the index $[G:H]$
is exponential --- for instance, if $G$ is an extension of $H$
by $\Z_p$ where $p$ is exponentially large --- then Beals' technique
takes exponential time.  

For a certain type of extension, we can construct circuits
analogous to Coppersmith's, which use quantum parallelism
to embed $\sigma$ in the $\rho$, sum over all $p$ cosets 
simultaneously, and apply the twiddle factor as well.  
Recall that $G$ is a {\em split extension} or {\em semidirect product} 
of $H$ by $T$, written $T \ltimes H$, if $H \lhd G$ and there is a 
transverse subgroup $T < G$ so that $T \cong G/H$.

\begin{definition}  Suppose $G$ is a split extension of $H$ by $T$, 
and let $S$ be a set of at most $\log_2 |T|$ generators for $T$, 
and suppose that $G$ is polynomially uniform with respect to a
tower of subgroups where $G=G_m$ and $H=G_{m-1}$ and a 
Bratteli diagram $\Brat$.  Then 
$G$ is a {\em homothetic extension} of $H$ by $T$ if
\begin{enumerate}
\item Given $\sigma \in \hat{H}$ and $\gamma \in S$, 
define $\sigma^\gamma(h) = \sigma(\gamma^{-1} h \gamma)$.  
Then for every $\sigma \in \hat{H}$, either $\sigma^\gamma = \sigma$, 
or the orbit of $q$ distinct representations 
$\sigma^{\gamma^j}$, for $0 \leq j < q$ where 
$q$ divides the order of $\gamma$, appears among the 
representations of $H$ given by $\Brat$.
\item For each $\gamma \in S$, 
there is a classical algorithm which runs in $\polylog(|G|)$ time
which, given a path $s$ in $\Brat$ indexing a row of $\sigma[s]$ and an
integer $j$, returns the size $q$ of $\sigma$'s orbit under conjugation
by $\gamma$, and returns a path $s^{\gamma^j}$ that indexes the same row of
$\sigma[s^{\gamma^j}] = \sigma^{\gamma^j}$.
\end{enumerate}
\end{definition}

\begin{theorem}
\label{thm:nicecopper}
If $G$ is a homothetic extension of $H$ by an Abelian group, 
then the Fourier transform of $G$ can be obtained
from the state~\eqref{eq:stateh}
using $\polylog(|G|)$ elementary
quantum operations . 
\end{theorem}

\noindent
{\em Proof.} It is easy to show that a homothetic extension of 
$H$ by $A \times B$ consists of a homothetic extension of
$H$ by $A$, followed by a homothetic extension by $B$.  
Therefore it suffices to prove the lemma for homothetic extensions 
by cyclic groups of prime power order, so
let $T$ be generated by $\gamma$ of order $p^z$. 

We recall some representation theory
from~\cite{Clifford:Representations,Rockmore:FastAbelian}.  Given $\sigma \in \hat{H}$, the
{\em stabilizer} of $\sigma$ is $K = \{x \in T: \sigma^x \cong \sigma\}$, 
and for a homothetic extension we can replace $\sigma^x \cong \sigma$
with $\sigma^x = \sigma$.  Then $K$ is the subgroup of $T$ of order $p^\ell$ 
generated by $\gamma^q$ where $q = p^{z-\ell}$, and $\sigma$'s orbit under 
conjugation by $\gamma$ is of size $q$.

The representations $\rho$ in which $\sigma$ appears can be
obtained in two steps.   First, we extend $\sigma$ to $K \ltimes H$ 
by multiplying $\sigma$ by one of the $p^\ell$ characters of $K$.  
This yields $\tau_b \in \widehat{K \ltimes H}$ where 
$\tau_b(\gamma^{qj} h) = \chi_b(j) \,\sigma(h)$ and
$\chi_b(\gamma^{qj}) = \omega_{p^\ell}^{bj}$. 
Since $d_{\tau_b} = d_\sigma$, we have $A_{\sigma,\tau_b} = 
\sqrt{1/p^\ell}$ and $\sigma$ embeds in a uniform superposition over 
the $\tau_b$, so we append a uniform superposition of
edges $1 \leq e \leq p^\ell$ where $b = e-1$.  
Combining this with the twiddle factor $\chi_b$ gives the unitary transformation
\begin{equation}
\label{eq:step1}
 \ket{\gamma^{qj+k}} \ket{s,t} 
  \to \ket{\gamma^k} \otimes \frac{1}{\sqrt{p^\ell}}
  \sum_{e=1}^{p^\ell} \omega_{p^\ell}^{(e-1)j} \ket{se,te}
\enspace .
\end{equation}  
Here we write the power of $\gamma$ in two registers $0 \leq j < p^\ell$
and $0 \leq k < q$.  Then this 
operation Fourier transforms the first register over $\Z_{p^\ell}$ 
and transfers the result to the $m$th register of $s$ and $t$.  
This transform can be carried out with $O(\log p^\ell \log \log p^\ell) =
O(\log |G| \log \log |G| )$ 
elementary operations~\cite{Hales:2000:IQF,Kitaev:Quantum}.  Note that $p^\ell$ takes
at most $\log |G|$ different values, and can be
obtained from the classical algorithm which computes $q$.

If $K = T$, then the $\rho \in \hat{G}$ containing $\sigma$ are simply 
the extensions $\tau_b$ and we're done.  If $K < T$, i.e.\ if $q > 1$, 
we carry out a second step as follows.  Each $\tau_b$ appears in a 
single induced representation $\rho_b$ whose restriction to $K \ltimes H$ 
is the direct product of all the representations in $\sigma$'s orbit, 
times $\chi_b$: that is,
$\res{\rho_b}{H} = \chi_b \oplus_{i=0}^{q-1} \sigma^{\gamma^i}$.  
The twiddle factor $\rho_b(\gamma^k)$ is then a permutation matrix which 
cycles these $p$ blocks $k$ times, with an additional phase change 
$\omega_{p^z}^{bk}$. This gives the unitary transformation
\begin{equation}
\label{eq:step2}
 \ket{\gamma^k} \ket{se,te}
  \to \omega_{p^z}^{(e-1)k} \ket{0} \ket{s^{\gamma^k}e, te} \enspace .
\end{equation}
Since $s^{\gamma^k}$ can be calculated by the classical algorithm
in $\polylog(|G|)$ time, and since it is easy to implement $\omega_{p^z}^{bk}$ 
with phase shifts $\omega_{p^z}^{2^y b}$ for $0 < y < \log_2 k$ 
conditioned on the binary digit sequence of $bk$, we can perform 
this operation in $\polylog(|G|)$ quantum steps.  
Composing~\eqref{eq:step1} 
and~\eqref{eq:step2} transforms the state~\eqref{eq:stateh} to the 
Fourier transform~\eqref{eq:stateg} over $G$.
\qed

\medskip \noindent \textbf{Closure under homothetic extensions 
and the metacyclic groups.} 
Theorem~\ref{thm:nicecopper} shows that the set of groups for which
circuits of $\polylog(|G|)$ size exist is closed under homothetic extensions
by Abelian groups.  It also generalizes the efficient quantum Fourier transform
of H{\o}yer~\cite{Hoyer:Efficient} for the {\em metacyclic} groups $\Z_q \ltimes \Z_p$, 
since these are homothetic extensions of $\Z_p$ by $\Z_q$.  
Note that the metacyclic groups include the dihedral groups (where $q=2$) 
and the affine groups (where $q=p-1$) as special cases.


\medskip \noindent \textbf{The general case.} 
In general, Abelian extensions can be slightly more complicated; 
consider extensions by $\Z_p$.
If $\sigma^\gamma$ is isomorphic to $\sigma$, rather than 
equal to it, $\gamma$ induces an additional twiddle factor 
$C(\gamma)$ which changes $\sigma$'s basis~\cite{Rockmore:FastAbelian}.  
This occurs, for instance, 
if $\gamma^p$ is an element of $H$ other than the identity, 
in which case the cyclic group generated by $\gamma$ is not 
transverse to $H$ and the extension is not split.  In this case 
$C(\gamma)$ is a $p$'th root of $\sigma(\gamma^p)$.  

\medskip \noindent \textbf{Relation to Coppersmith's  circuit.}
Let $\gamma$ be a generator of $G=\Z_{2^n}$.  Then
$G$ is an extension of $H=\Z_{2^{n-1}}$ with transversal
$\{1,\gamma\}$.  Since $\gamma^2 \neq 1$, $\gamma$ 
induces an additional phase shift 
$C(\gamma) = \sqrt{\chi_b(\gamma^2)} = \omega_{2^n}^b$.
(Similarly, the additional phase shift in~\eqref{eq:step2} is due to the 
fact that $\Z_{p^z}$ is not a split extension of $\Z_{p^\ell}$.)
In Coppersmith's circuit, $C(\gamma)$ appears as a set of phase shift 
gates conditional on the low-order bit of $j$.  Finally, the 
Hadamard gate in Coppersmith's circuit is precisely the 
operation~\eqref{eq:step1} in the case $p=2$, $\ell = 1$ and $q=1$, 
and where we use the same qubit register for $e$ (the high-order bit
of the frequency) as for $\alpha$ (the low-order bit of the time).

\medskip \noindent \textbf{The quaternionic groups.} 
Another example is the generalized quaternion group, which is an
extension of $H=\Z_{2n}$ by $\Z_2$ where $\gamma^2$ is the element
of order 2 in $H$.  Then $C(\gamma) = \sqrt{\sigma(\gamma^2)} = 1$ or $i$.  
P{\"u}schel, R{\"o}tteler and Beth~\cite{PuschelRB99} gave an efficient quantum 
Fourier transform for these groups in the case where $n$ is a power of 2.  
Of course, these groups are extensions of Abelian groups with bounded 
index, so Lemma~\ref{lem:polyindex} already provides an 
efficient QFT for them.

\medskip \noindent \textbf{Metabelian groups.} 
Even if an extension is neither homothetic nor of polynomial index, we 
can still construct an efficient QFT if we can apply arbitrary powers of 
$C(\gamma)$ in polynomial time.  This is true, for instance, if $C(\gamma)$ 
is of polynomial size, which is true whenever all the representations 
of $H$ are of polynomial size.  This includes the {\em metabelian} groups, 
i.e.\ split extensions of Abelian groups by Abelian groups, since all the
representations of $H$ are one-dimensional.  
We discuss this further in the full paper.

\section{Conclusion and open problems}
\label{sec:Conclusion}

The separation of variables is in essence a coarse scale use of a
factorization of the dual, using blockwise redundancy as well as
sparseness. It is possible to use the Bratteli diagram indexing and
accompanying path factorizations in a more precise fashion,
effectively looking for redundancy and sparsity on the level of
individual elements. This finer analysis is responsible for the
fastest known classical FFTs for the groups $\SL_2(q)$, as well as 
$S_n$ and its wreath products~\cite{MaslenR:Cooley}.  It would
be interesting to investigate the possibility of adapting these
techniques to the quantum setting.

\section*{Acknowledgements}

The authors gratefully acknowledge the support of the National Science
Foundation under grants CCR-0220264, EIA-0218443, and CCR-0093065 and
the hospitality of the Mathematical Sciences Research Institute, where
portions of this research were completed.

\bibliography{qfft-refs,stoc1990,stoc2000,focs2000,soda,sicomp}
\end{document}